\newif\ifTwoColumn%
\newif\ifSUBMIT%
\newif\ifCOMMENTS%
\newif\ifFIGs%
\newif\ifFIGoneColumn%
\let\ifTwoColumn\iftrue%
\let\ifCOMMENTS\iftrue%
\let\ifSUBMIT\iffalse%
\let\ifFIGs\iftrue%
\let\ifFIGoneColumn\iftrue%
\ifTwoColumn%
  \documentclass[aps, pre, reprint, amsmath, superscriptaddress, floatfix]{revtex4-1}
\else
  \documentclass{revtex-4.1}
\fi
\usepackage{xparse}
\usepackage{graphicx}
\usepackage{amsmath}
\usepackage{xcolor}
\usepackage{hyperref}
\usepackage[utf8]{inputenc}
\usepackage{enumitem}
\usepackage{acronym}

\usepackage{todonotes}

\usepackage{array,mathtools,amssymb,booktabs}

\usepackage[normalem]{ulem}
\definecolor{edits}{RGB}{200,0,0}
\definecolor{strike}{RGB}{130,0,0}

\definecolor{mygray}{RGB}{128,128,128}

\newcommand{\eg}{e.\,g.}

\newcommand{\ie}{i.\,e.}

\newcommand{\sno}[1]{_\mathrm{#1}}
\newcommand{\no}[1]{\mathrm{#1}}

\renewcommand{\mathbf}[1]{\boldsymbol{#1}}

\begin{document}

\title{Thermal decay rates of an activated complex
in a driven model chemical reaction}

\author{Robin Bardakcioglu}
\author{Johannes Reiff}
\author{Matthias Feldmaier}
\author{J\"org Main}
\affiliation{%
Institut f\"ur Theoretische Physik I,
Universit\"at Stuttgart,
70550 Stuttgart,
Germany}
\author{Rigoberto Hernandez}
\email[Correspondence to: ]{r.hernandez@jhu.edu}
\affiliation{
    Department of Chemistry,
    Johns Hopkins University,
    Baltimore, Maryland 21218, United States
}
\affiliation{
    Departments of Chemical \& Biomolecular Engineering,
    and Materials Science and Engineering,
    Johns Hopkins University,
    Baltimore, Maryland 21218, United States
}
\date{\today}

\begin{abstract}
Recent work has shown that in a non-thermal, multidimensional system, the
trajectories in the activated complex possess different instantaneous and
time-averaged reactant decay rates.
Under dissipative dynamics, it is known that these trajectories,
which are bound on the normally hyperbolic invariant manifold (NHIM),
converge to a single trajectory over time.
By subjecting these dissipative systems to thermal noise, we
find fluctuations in the saddle-bound trajectories and their
instantaneous decay rates.
Averaging over these instantaneous rates results in
the decay rate of the activated complex in a thermal system.
We find, that the temperature dependence of the
activated complex decay in a thermal system can be linked to
the distribution of the phase space resolved decay rates on the NHIM
in the non-dissipative case.
By adjusting the external driving of the reaction,
we show that it is possible to influence
how the decay rate of the activated complex changes with rising temperature.
\end{abstract}

\maketitle

\acrodef{NN}{neural network}
\acrodef{BCM}{binary contraction method}
\acrodef{DS}{dividing surface}
\acrodef{TST}{transition state theory}
\acrodef{TS}{transition state}
\acrodef{LD}{Lagrangian descriptor}
\acrodef{LMA}{local manifold analysis}
\acrodef{NHIM}{normally hyperbolic invariant manifold}
\acrodef{PSOS}{Poincar\'e surface of section}
\acrodef{EOM}{equation of motion}

\section{Introduction}

In order to predict the rate of chemical reactions,
\ac{TST}~\cite{eyring35, wigner37,pitzer,pech81,truh79,truh85,%
  truhlar91,truh96,truh2000,KomatsuzakiBerry01a,Waalkens2008,hern08d,%
  Komatsuzaki2010,hern10a,Henkelman2016}
utilizes a \ac{DS} in phase space~\cite{keck67,peters14a}
to determine when reactants decay into products under
quasi-equilibrium conditions~\cite{mill98}.
We and others~\cite{hern08d,hern17h,hern19a,Waalkens2008,hern10a,Komatsuzaki2010,
Henkelman2016,pollak90a,Uzer02,KomatsuzakiBerry01a,dawn05a,wiggins16}
have extended the use of \ac{TST} to address reactions far out of
equilibrium leading to rates, resolution of the reactive geometry,
or the reaction paths.
Such conditions can arise when the reactants
respond to external stimuli---\eg\ under driven conditions
or collective effects of the reacting environment.

The accuracy of the \ac{TST} rate depends on the accuracy of the \ac{DS}
to truly divide the space between reactants and products.
That is, it must satisfy the condition that no reacting particle
crosses it more than once.
Unfortunately, in a driven system, it is often not enough to classify
molecules based only by their spatial configuration in large part
because the structure of the reaction geometry is itself time-dependent.
The \ac{DS} must then be extended to the full phase space of the
system in a time-dependent frame to ensure that it is free of
recrossings~\cite{hern16h,hern17g,hern17h,hern19a}.

An additional complexity arises in activated processes which must either
explicitly address a solvent by extending the number of degrees to a
macroscopic degree---\eg\ moles---or implicitly by including them through
a formalism such as that of Brownian motion~\cite{brown28,kram40,rmp90}.
In either case, the coupling between the reactants and the solvent can
be surmised through an effective friction.
Activated processes have been seen to undergo a Kramers turnover~\cite{kram40,rmp90}
in the rate as they are solvated from low to high friction~%
\cite{mm86,pgh89,hern08g,hern12e,hern16c},
and hence the exact value of the friction is important in determining the rate.
In general, the activated complex is a collection of unstable configurations
located near the energy barrier between the reactants and products.
In the modern language of differential geometry, this has become associated
with the so-called
\ac{NHIM}~\cite{Lichtenberg82,hern93b,hern93c,Ott2002a,wiggins2013normally,hern19e}.
It is a co-dimension 2 manifold in the phase space of the system,
which is characterized by the condition that trajectories started on
that manifold stay there when propagated forward or backward in time.

The \ac{DS} (or \ac{NHIM}) is complementary to the brute force calculation of
reaction rates using trajectories that start from the reactant region.
Such trajectories only count if they cross to products and are rare
in activated processes.
Nevertheless, the few that are reactive, cross an exact \ac{DS} once
and only once.
Avoiding the work of determining the nonreactive trajectories from the
reactant region, one thus typically focuses on the rate of trajectories
leaving from the \ac{DS} which when exact---because of the non-recrossing
condition---gives the flux-over-population
rate~\cite{lang69,Pollak95c,farkas27,pollak05a}.
Either because of time dependence or dimensionality,
the NHIM itself can generally
accommodate a set of trajectories that neither enter nor leave it.
In recent work, we have explored the stability of this class of trajectories
as we have conjectured that their decay is connected to the decay of
the reactive trajectories~\cite{hern19a, hern19e, hern20d}.
In dissipative driven systems,
due to the properties of the \ac{NHIM},
trajectories within it
converge towards a single \ac{TS} trajectory after a sufficiently long time in the
saddle region.
Those trajectories near it, will also be trapped
towards a single trajectory of the \ac{NHIM}~%
\cite{dawn05a,dawn05b,hern08d}.
Obtaining the decay rate of the trajectories within the
\ac{NHIM},
and a determination of how the thermal environment affects them
is the primary contribution of this work.

Using the system and methods described in Sec.~\ref{sec:SystemMethods}, we can
use geometric structure obtained directly
to determine rates
in driven chemical reactions that are not isolated,
but rather coupled to a dissipative environment.
This is a necessary advance for the use of the
nonrecrossing dividing
surfaces---\emph{viz.}\ the time-dependent \ac{DS} attached to the \ac{NHIM}---that
we and others~\cite{dawn05a,dawn05b,hern08d,Bartsch12,hern19a,hern19e,hern20d}
have been developing
because many chemical reactions of interest occur in a solvent.
The results presented in Sec.~\ref{sec:results}
provide a demonstration of the stochastic time-dependent motion of the
\ac{NHIM} at fixed orthogonal modes (Sec.~\ref{sec:erratic}) and the
collapse of the transition states under dissipation towards a single
trajectory on the  \ac{NHIM} (Sec.~\ref{sec:dissipative}).
The time dependence of the
instantaneous reactant decay rate and the temperature dependence
of the average decay rate of the activated complex over long times
are shown in Sec.~\ref{sec:thermal}.
The temperature-dependent behavior of the average decay
rate is linked to the phase space resolved average decay rate of the
non-thermal system.
We also find that
the temperature dependence of the decay rate can be influenced
by changing the oscillation of the periodic driving.

\section{System and methods}\label{sec:SystemMethods}

In this section, we first recapitulate the representation of a
chemical reaction~\cite{hern19T3, hern19e, hern19a}.
As in earlier work, we impose Langevin dynamics to represent the
influence of the bath, and use a driven saddle potential to reveal the
decay rates of trajectories within the \ac{NHIM}.
The specifics of the system and the associated \ac{EOM} are summarized
in Sec.~\ref{sec:model}.
The unstable transition states, \ie, the trajectories on the \ac{NHIM},
can then be constructed using the approach described in
Sec.~\ref{sec:projection}~\cite{hern18g,hern20d}.
The instantaneous decay rates at coordinates of the \ac{NHIM}
and the average decay rates along a transition state
can then be constructed as summarized in Sec.~\ref{sec:rate_methods}.

\subsection{Model chemical reaction}
\label{sec:model}
The dynamics of the system under investigation is given by the Langevin
equation,
\begin{subequations}\label{eq:langevin}
\begin{align}
    \dot{\boldsymbol{v}} &=
    - \gamma \boldsymbol{v}
    + \boldsymbol{\xi} (t)
    - \boldsymbol{\nabla} V(\boldsymbol{x}, t)\,,\\
    \dot{\boldsymbol{x}} &= \boldsymbol{v}\,,
\end{align}
\end{subequations}
where $\boldsymbol{x}$ is the coordinate vector of the system,
$\boldsymbol{v}$ the corresponding velocities, $t$ the time, and
$\gamma$ the friction coefficient.
The vector $\boldsymbol{\xi}(t)$ represents the fluctuations around
the time-dependent mean-force potential
$V(\boldsymbol{x},t)$.
Here, we represent each component as white noise which
satisfies the fluctuation-dissipation theorem~\cite{kubo66, keizer76, hern99a}
with respect to the specified friction,
\begin{align}
    \left< \xi_i(t) \right> &= 0\,, \label{eq:noise-zero}\\
    \left< \xi_i(t)\xi_j(t^\prime) \right> &=
    2\gamma T \delta_{ij}\delta(t-t^\prime)\,,\label{eq:noise}
\end{align}
where $\delta_{ij}$ and $\delta(t - t^\prime)$ represent the Kronecker delta and
Dirac delta distribution, respectively.
The temperature $T$ is given in units
where the Boltzmann constant $k_B$ is $1$
so as to give it the same units as energy.
Note that the same noise sequence is used for all trajectories.

The specific potential investigated here has been used many times before in
previous work~\cite{hern17h, hern18c, hern18g, hern19a, hern19e, hern20d,%
  hern19T3, hern20n}.
It is a two-dimensional rank-1 saddle potential of the form
\begin{align}
    V(x, y, t) &= 2 \exp{
        \left(-{\left[ x - 0.4 \sin{(\omega t)}\right]}^2\right)
    } \nonumber \\
    &+ 2 {\left(y - \frac{2}{\pi}
    \arctan{(2x)}\right)}^2\,.
    \label{eq:potential}
\end{align}
This potential models a reaction over an energy barrier
oscillating with frequency $\omega$
and provides a nonlinear coupling between reaction coordinate and
orthogonal mode along the reaction path.
The coordinate component $x$ approximates the reaction coordinate,
\ie, the unstable direction of the saddle potential, by construction.
Likewise, the $y$ component approximates the orthogonal modes.
These coordinates $\boldsymbol{x} = (x,y)$, together with their velocities
$\dot x = v_x,\ \dot y = v_y$ form the phase space of the system.

\begin{figure}
    \centering
    \includegraphics[width=\linewidth]{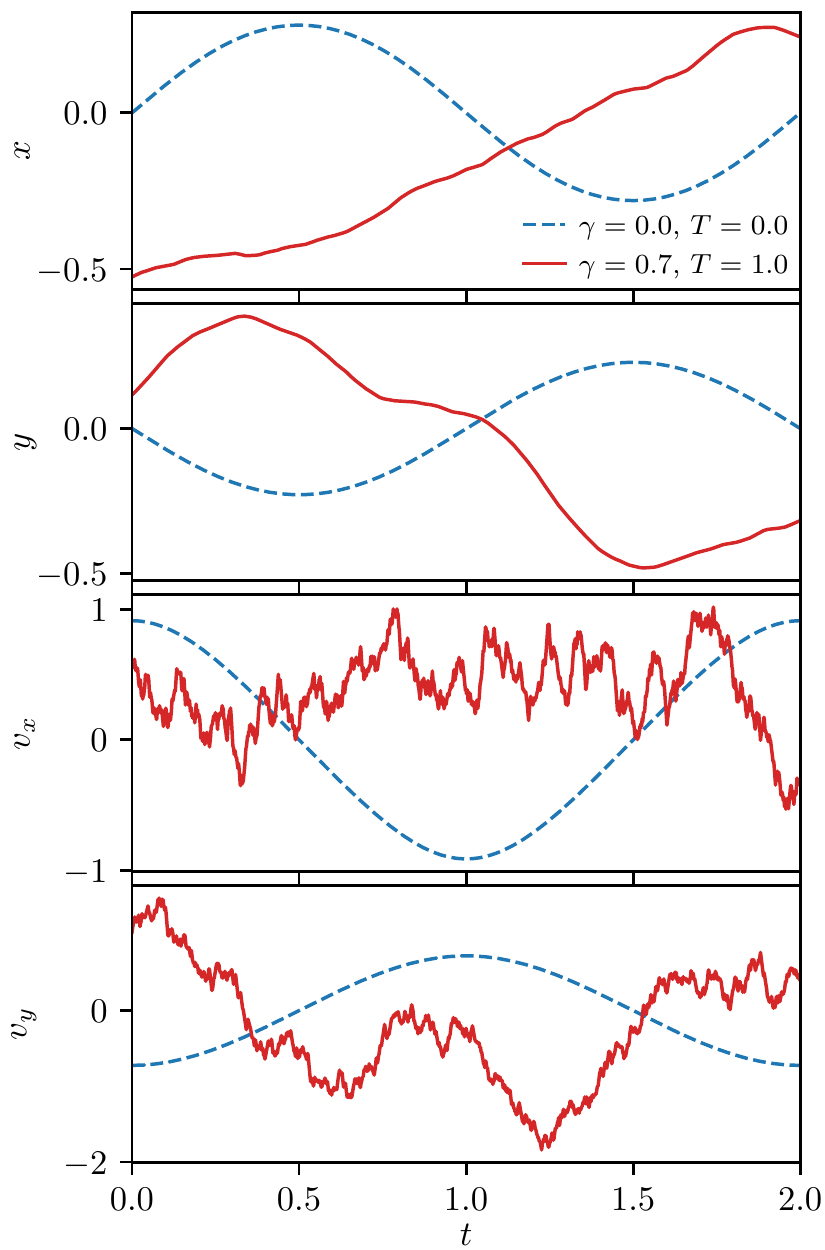}
    \caption{%
        Phase space coordinates $(x, y, v_x, v_y)$
        of two trajectories arbitrarily selected
        from the ensemble of \acs{TS} trajectories on the \acs{NHIM}
        at $T = 0$ and $T = 1.0$, respectively.
        The non-thermal trajectory ($T = 0$, $\gamma = 0$)
        is periodic with the driving frequency $\omega = \pi$
        of the potential, as can be readily seen.
        The thermal trajectory ($T = 1.0$, $\gamma = 0.7$), in contrast,
        exhibits fluctuations characteristic of the stochastic driving,
        and it spreads out of sync with the periodic driving.}
    \label{fig:trajectories}
\end{figure}

The effect of the Langevin dynamics on the trajectories of the system is
illustrated in Fig.~\ref{fig:trajectories}.
The phase space coordinates of two arbitrary
\ac{TS} trajectories on the \ac{NHIM} are shown.
In the non-thermal case we can see that
the trajectory follows a smooth path in phase space and it
is periodic in sync with the deterministic driving.
In contrast, the trajectory
of the thermal system fluctuates significantly, especially in the velocities.
The stochastic and aperiodic motion of the thermal trajectories
presents a new challenge for
our earlier methods on non-thermal
systems~\cite{hern18g, hern19a, hern19e, hern20d} addressed here.

\subsection{Identification of the \ac{NHIM}}
\label{sec:manifolds}

The \ac{NHIM}, barring any general definition and here only limited to a rank-1
saddle potential, is the set of all trajectories in phase space bound forever
to the saddle region in both forward and backward in time directions.
It is a manifold of co-dimension two in phase space.
It is also associated with
a pair of stable and unstable manifolds of co-dimension one, whose
closures intersect at the \ac{NHIM}.
For our model chemical reaction [Eq.~\eqref{eq:potential}],
the position $(x, v_x)^{\no{NHIM}}(y, v_y)$ of the
\ac{NHIM} can be parameterized as a function of the stable orthogonal modes
$y$ and $v_y$ at a specific time $t$.

\begin{figure}
    \includegraphics[width=\linewidth]{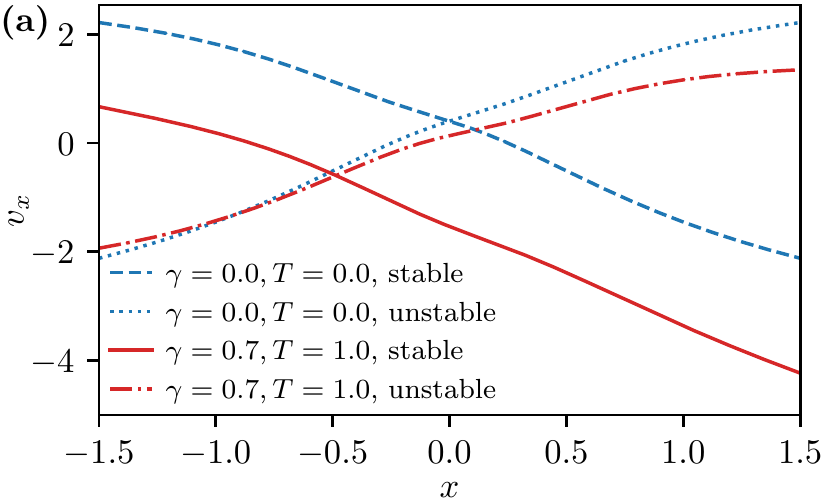}
    \includegraphics[width=\linewidth]{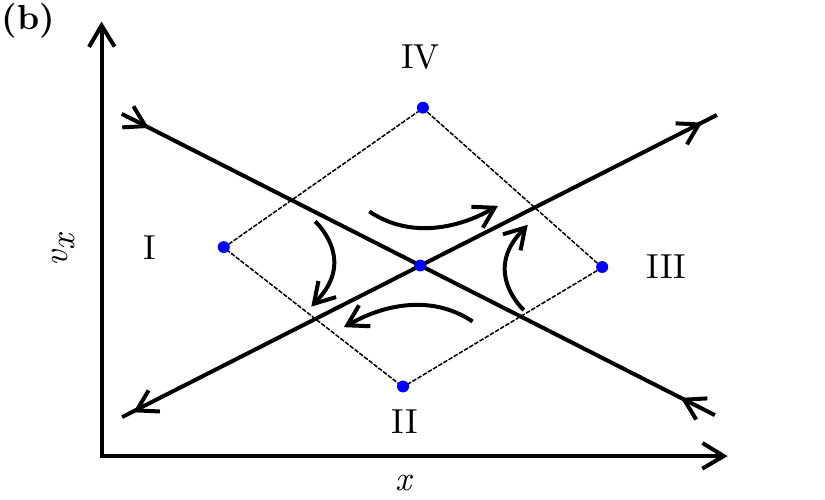}
    \caption{%
        (a) Stable and unstable manifold at orthogonal modes $y = v_y = 0$
        for a non-thermal and thermal system
        with driving frequency $\omega = \pi$.
        The intersection of these manifolds mark the $x$ and $v_x$ coordinate
        of the \acs{NHIM} for these orthogonal modes.
        Crosses such as these can be obtained
        for any orthogonal modes $(y, v_y)$ and time $t$.
        Thus, one can express the reaction coordinates
        $(x, v_x)^\mathrm{NHIM}(y, v_y, t)$
        of the \acs{NHIM} as functions of initial time and orthogonal modes.
        (b) Sketch of the stable and unstable manifold.
        The arrowheads indicate the projected path trajectories take
        within the four separated regions.
        A quadrangle with a vertex in each of the four regions
        illustrates the \acl{BCM},
        an iterative routine to find the \acs{NHIM}
        at the intersection of the manifolds, see Sec.~\ref{sec:manifolds}.}
    \label{fig:manifolds}
\end{figure}

The motion of individual trajectories
in a close neighborhood of the \ac{NHIM} is
stochastic for a thermal system as illustrated in Fig.~\ref{fig:trajectories}.
Nevertheless, the corresponding stable and unstable manifolds
in Fig.~\ref{fig:manifolds}(a) remain smooth, and generally so.
Thus, even for a thermal system, the phase space
in a local neighborhood looks similar to a non-thermal system,
as also observed in Refs.~\cite{hern18g,hern19a,hern19e,hern20d}.
Hence, the typical cross-like intersection of the stable
and unstable manifolds is preserved and
the position of the \ac{NHIM}
can be directly obtained from the intersection of the stable and the unstable manifolds.
In a thermal system, however, the intersection $(x, v_x)^{\no{NHIM}}(y, v_y)$
will not only depend on the specific choice of the orthogonal modes $(y, v_y)$,
but also on the parameters $\gamma$ and $T$.
This finding is illustrated in Fig.~\ref{fig:manifolds}(a) for the paradigmatic
system used throughout this work.

The stable and unstable manifolds have the characteristic property that any
trajectory near them is propagated towards or away from the \ac{NHIM},
respectively.
Only at the intersection of their
closures---\emph{viz.}\ the \ac{NHIM}---
trajectories are unstably bound
to the saddle region.
As the stable and unstable manifold are themselves invariant manifolds,
they cannot be crossed by any trajectory of the system.
This effectively separates the phase space
in the close neighborhood of the \ac{NHIM}
into four distinct regions demarcated by these manifolds,
see Fig.~\ref{fig:manifolds}(b).
For a chosen
saddle region $x\in[x\sno{min}, x\sno{max}]$ with appropriately
chosen boundaries $x\sno{min}$ and $x\sno{max}$,
the dynamics of trajectories is characteristic for any of these regions, as
indicated with the black arrows in Fig.~\ref{fig:manifolds}(b).
In region (I), non-reactive trajectories originate from the reactant side
$x < x\sno{min}$ in the past and fall back to the reactant side in the future.
The same is true for region (III) which contains all trajectories
that originate from the product side $x>x\sno{max}$ and also fall back to the product side.
Regions (II) and (IV) hold the reactive trajectories from the reactant to the product side
or, respectively, vice versa.

We use the \ac{BCM}~\cite{hern18g}
to numerically find the \ac{NHIM}
at a given time $t$ for a given set $(y, v_y)$ of orthogonal modes.
The procedure of the \ac{BCM} is initiated
with a quadrangle having its four vertices (blue dots
in Fig.~\ref{fig:manifolds}(b)) in each of the four reactive and non-reactive
regions in a close neighborhood of the \ac{NHIM}.
In successive steps, the midpoint between
pairs of nearby points of the quadrangle is propagated to determine
which region it belongs to and then replaces the corresponding
point.
The area of the quadrangle thus shrinks and its center converges
to the intersection $(x, v_x)^{\no{NHIM}}(y, v_y)$ of the manifold.
The resulting $(y, v_y)$ corresponds to the
position $(x, v_x)$ of the \ac{NHIM}.
This convergence is exponentially fast and, therefore,
the \ac{BCM} is very efficient in finding the \ac{NHIM}
for a given set of orthogonal modes.
For the technical details of the \ac{BCM}, we refer the reader
to Ref.~\cite{hern18g}.

\subsection{Trajectories on the \ac{NHIM}}
\label{sec:projection}

Due to the hyperbolic nature of the \ac{NHIM}, trajectories on the
\ac{NHIM} are unstable.
Small deviations from the \ac{NHIM} will grow
exponentially in time until the trajectory leaves its immediate vicinity.
Thus any deviation in a point relative to the \ac{NHIM}, no matter how small,
will lead its subsequent propagation to fall off of it eventually.
This presents a challenge to the propagation of trajectories on the \ac{NHIM}
using numerical simulations.

These numerical deviations can be suppressed through
a machine-learning representation of the \ac{NHIM} projecting
them back onto the \ac{NHIM} as has been done
in Ref.~\cite{hern20d}.
Here, we employ this approach leveraging the numerically accurate
\ac{BCM} presented in Ref.~\cite{hern18g}.
Given the system and bath parameters,
the reaction coordinate and corresponding velocity
of the \ac{NHIM} can be integrated in time
as a function of the remaining orthogonal modes in the usual way.
By subsequently projecting the trajectory back onto the \ac{NHIM} it is
effectively propagated in a system with reduced dimensions which
is spanned by the orthogonal modes.
The result is a trajectory that remains on the \ac{NHIM}.

\subsection{Decay rates of reactant population}
\label{sec:rate_methods}

The relative stability of the \ac{NHIM} can be characterized through
the instantaneous decay rates $k(y, v_y, t)$
of the reactant population
for the specific orthogonal modes $(y, v_y)$ in its close neighborhood
at each time $t$.
These decay rates can be obtained
directly by propagating ensembles of reactive trajectories
but this can become cumbersome and numerically expensive.
The \ac{LMA} was introduced in Ref.~\cite{hern19e} to overcome
this problem by
leveraging the linear dynamics near the \ac{NHIM}.

Since the dynamics relative to the transition state is linear, it is possible
to propagate trajectories using
a linear map, the fundamental matrix $\boldsymbol{M}$, obtained via
\begin{equation}
    \dot{\boldsymbol{M}} = \boldsymbol{J}(t)\boldsymbol{M}\, ,
    \label{eq:fundamental}
\end{equation}
with the initial condition of $\boldsymbol{M}(t_0) = \mathbf{1}$, and a Jacobian
\begin{equation}
	\boldsymbol{J}(t) = \frac{\partial(\dot{\boldsymbol{x}}, \dot{\boldsymbol{v}})}{\partial(\boldsymbol{x}, \boldsymbol{v})}
               =
               \begin{pmatrix}
                0 & 0 & 1 & 0 \\
                0 & 0 & 0 & 1 \\
                - \frac{\partial^2V}{\partial x^2} & - \frac{\partial^2V}{\partial x\partial y} & -\gamma & 0 \\
                - \frac{\partial^2V}{\partial x\partial y} & - \frac{\partial^2V}{\partial y^2} & 0 & -\gamma\\
               \end{pmatrix}\,
    \label{eq:jacobian}
\end{equation}
parameterized in time along a specific trajectory of
the transition state.
The stochastic force $\boldsymbol{\xi}(t)$ does not contribute to any component
of
the Jacobian in Eq.~\eqref{eq:jacobian} as it is purely time-dependent.
However, this does not mean that it is neglected in the Langevin dynamics.
The Jacobian~\eqref{eq:jacobian} describes the motion relative
to a transition state whose
trajectory is fluctuating under the influence of the stochastic force.
That means that although the dynamics relative to the close vicinity of the
transition state might not be fluctuating, the total dynamics still does.
Using the linear dynamics near the transition state according to
Eq.~\eqref{eq:fundamental}, we can extract the instantaneous motion of a
particle ensemble near that state from the Jacobian.

The \ac{LMA} models the presence of such a uniform particle ensemble in the close neighborhood
of the \ac{NHIM}.
For the system of Eqs.~\eqref{eq:langevin} and \eqref{eq:potential}
at specific orthogonal modes $(y, v_y)$ and time $t$,
we can obtain a local instantaneous decay rate
\begin{equation}
    k(y, v_y, t) =   \frac{\Delta v^\mathrm{u}_x}{\Delta x^\mathrm{u}}
                     (y, v_y, t)
                   - \frac{\Delta v^\mathrm{s}_x}{\Delta x^\mathrm{s}}
                   (y, v_y, t)\, ,
    \label{eq:lma}
\end{equation}
where $\Delta v^\mathrm{s,u}_x/\Delta x^\mathrm{s,u}$ represents the slopes of
the stable and unstable
manifolds in the corresponding $(x, v_x)$ cross section
of the full phase space near the
transition state.
Determining these slopes, or rather the instantaneous decay rate as the
difference of these slopes, is the main objective of the numerical
implementation of the \ac{LMA}.
A derivation of the \ac{LMA}
can be found in Ref.~\cite{hern19e},
and the additional corrections
that would be necessary for more general cases
can be found in the Supplementary Material of Ref.~\cite{hern20m}.

An average rate for the decay relative to trajectories on the NHIM
is obtained by computing the time average of
the instantaneous decay rates
for a sufficiently long time to obtain convergence.
For a trajectory that is initialized at time $t_0$ and position
$(y_0, {v_y}_0)$ on the \ac{NHIM},
and parameterized by the orthogonal modes $y(t)$ and $v_y(t)$,
this average yields
\begin{equation}
	\bar k(y_0, {v_y}_0, t_0) =
              \lim_{\tau\rightarrow\infty}
                       \frac{1}{\tau} \int_{t_0}^{t_0 + \tau}
	k(y(t^\prime), v_y(t^\prime), t^\prime)\,\mathrm{d}t^\prime\,.
    \label{eq:average}
\end{equation}
In the special case of $T=0$, we find that the trajectories are
periodic or quasi-periodic,
and it suffices to integrate for
the period or quasi-period.
An alternative approach to obtain mean decay rates is provided by
a Floquet analysis of said trajectory~\cite{hern19e, hern14f}
\begin{equation}
    \bar k(y_0, {v_y}_0, t_0) =
           \lim_{\tau\rightarrow \infty}
              \frac{1}{\tau} \left(\ln{|m_\mathrm{l}(\tau)|}
                - \ln{|m_\mathrm{s}(\tau)|} \right)\, ,
    \label{eq:floquet-rate}
\end{equation}
where $m_{\mathrm{l,s}}(t)$ are the eigenvalues of
the fundamental matrix $\boldsymbol{M}$.
It reduces to the monodromy matrix when the trajectory is periodic.
Here, the subscripts l and s denote the eigenvalues with the largest and smallest
absolute values, respectively~\cite{hern14f, hern19e}.

\section{Results and discussion}
\label{sec:results}
\subsection{Stochastic motion of the NHIM under noise}
\label{sec:erratic}

\begin{figure}
    \includegraphics[width=\linewidth]{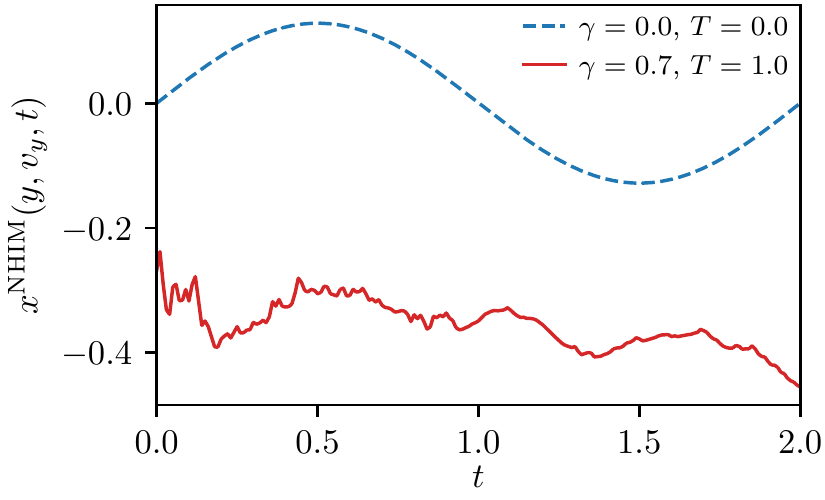}
    \caption{%
        Position $x^\mathrm{NHIM}$ of the \acs{NHIM}
        for fixed $y = v_y = 0$ and $\omega = \pi$
        as a function of time.
        Without the Langevin terms,
        the \acs{NHIM} (dashed curve) oscillates periodically,
        as one would expect from the periodically driven saddle potential.
        With the Langevin terms, however, the \acs{NHIM} (solid curve)
        moves stochastically.
        Compared to a
        trajectory on the \acs{NHIM} for the same parameters, the reaction
        coordinates $x$  does not fluctuate as
        much as the \acs{NHIM} in phase space,
        see Fig.~\ref{fig:trajectories}.
    }
    \label{fig:erratic}
\end{figure}

The time-dependent \ac{NHIM} and the associated
decay rates in a driven chemical reaction
are resolved here
for the model system of Eq.~\eqref{eq:langevin}.
If the dissipative and stochastic terms are excluded,
the result is a smooth oscillating motion
with the same period as that of the oscillating potential~\cite{hern20m}.
However, when the system is subject to
the Langevin terms---\emph{viz.}\ friction and thermal noise---the
motion of the \ac{NHIM} becomes stochastic as illustrated in
Fig.~\ref{fig:erratic}.
Here, the expected behavior
in the time-dependence of $x^\mathrm{NHIM}$
for a specific set of coordinates $(y, v_y)$
in the dynamics without and with the Langevin terms
is apparent.
The latter case now exhibits stochastic fluctuations
in both position (as shown) and momentum (as not shown)
space.
They are in response to the combination of
the collective stochastic thermal driving and the
periodic driving terms.
Such fluctuations were not seen in the position space
of the \ac{TS} trajectories shown in
Fig.~\ref{fig:trajectories} because in this relatively
weak friction regime, the high-frequency
fluctuations are very small.
However, in Fig.~\ref{fig:erratic},
the \ac{NHIM} does exhibit
short-time fluctuations in the position space
as a manifestation of the overall phase space
motion.
Nevertheless, the \ac{DS} is recrossing-free as it
incorporates the noise.

\subsection{Dissipative dynamics on the NHIM}
\label{sec:dissipative}

\begin{figure}
    \includegraphics[width=\linewidth]{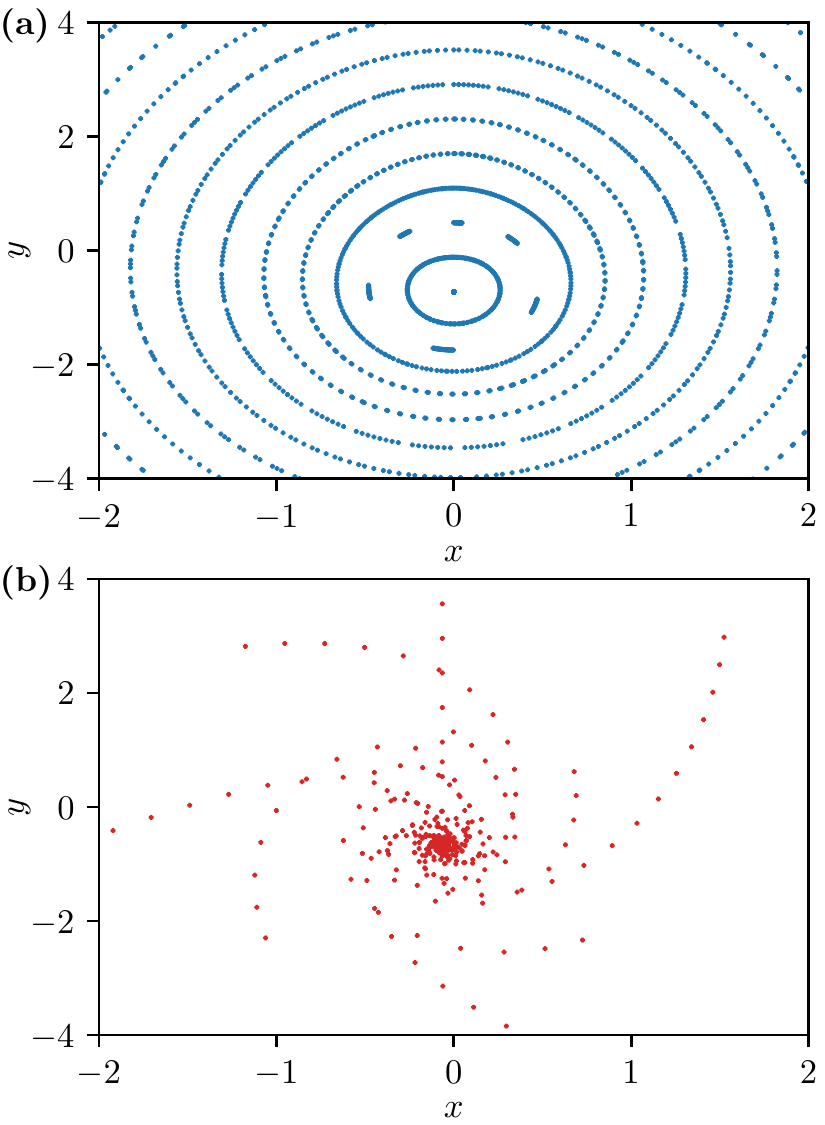}
    \caption{%
        The \acsp{PSOS} of the system defined by
        Eqs.~\eqref{eq:langevin} and \eqref{eq:potential} for varying friction
        at temperature $T = 0$ and saddle frequency  $\omega = \pi$.
        The non-thermal system with friction $\gamma = 0$ and
        the thermal system with  $\gamma = 0.2$ are shown in
        panels (a) and (b), respectively.
        In panel (a), the regularity of the system is demonstrated
        by the elliptic fixed point at the center and the surrounding tori.
        In panel (b), for trajectories with similar initial conditions as in (a),
        the stroboscopic dynamics on the \acs{NHIM} collapses to a
        fixed point in the dissipative case.
    }
    \label{fig:psos}
\end{figure}

Although we have seen that the dynamics of the system in the general
case of Eq.~\eqref{eq:langevin} for the potential in \eqref{eq:potential}
is not periodic, it is nevertheless instructive to examine the
stroboscopic \ac{PSOS} of its trajectories.
Similar to the approach used in
Refs.~\cite{hern19e, hern19T3}, we record the position of trajectories in the
$(y, v_y)$ section in phase space
at time steps equal to the period of the driving.
As we start with a point on the \ac{NHIM},
it necessarily must stay on the \ac{NHIM}.
Thus, the coordinates in the \ac{PSOS},
as seen \eg\ in Fig.~\ref{fig:psos},
remain projected onto the \ac{NHIM}.

The \acp{PSOS} of Fig.~\ref{fig:psos} show the contrast in
the dynamics upon the introduction of friction.
In the upper panel, the dynamics on the \ac{NHIM} is regular.
It gives rise to the expected stable concentric tori
and an elliptic fixed point.
In the bottom panel, as a result of the friction,
the would-be tori now spiral
towards a fixed point in the stroboscopic projection.
This fixed point refers to a time-periodic trajectory in phase space,
which acts as an attractor for particles on the \ac{NHIM}.

\subsection{Thermal decay rates}
\label{sec:thermal}

\begin{figure}
    \includegraphics[width=\linewidth]{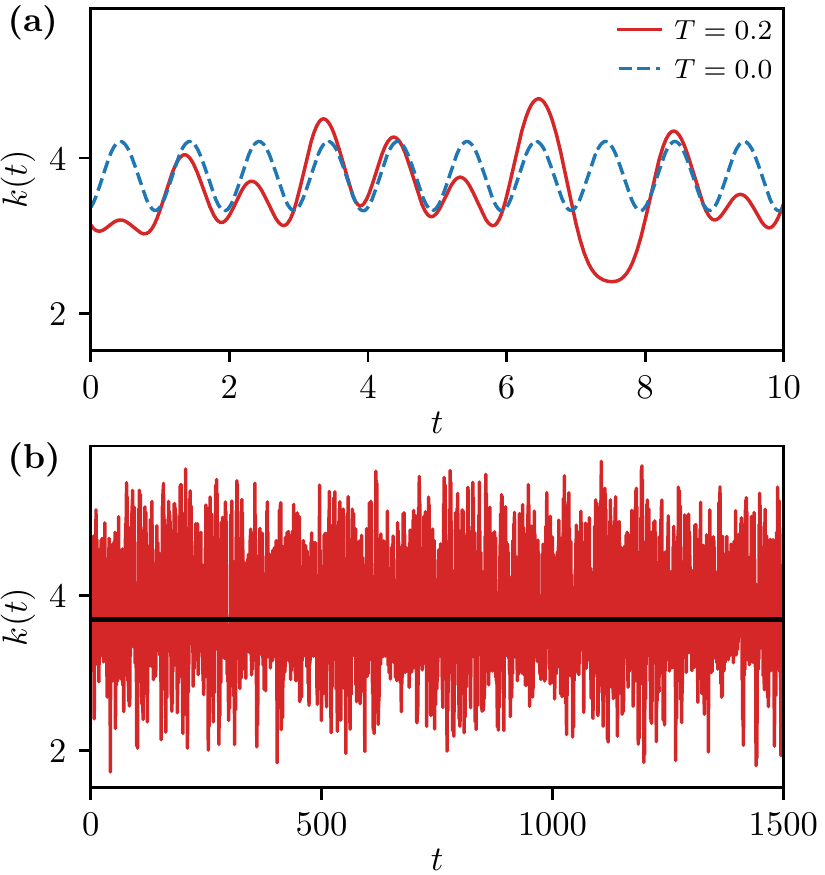}
    \caption{%
        (a) Comparison of instantaneous decay rates of
        the equilibrium trajectory on the \acs{NHIM}
        as defined in Sec.~\ref{sec:thermal}
        for friction $\gamma = 0.2$ and temperatures $T = 0.2$ and $0$,
        respectively.
        The driving frequency in both cases is $\omega = \pi$.
        The instantaneous rate of the thermal complex roughly follows the
        oscillations of the non-thermal trajectory but exhibits fluctuations to a
        certain degree.
        (b) Instantaneous rate of the thermal activated complex in
        over a longer time
        interval.
        Stochastic fluctuations of the rate become more evident over hundreds
        of saddle oscillation periods. The
        dashed line highlights the time-average of the rate.}
    \label{fig:rate_comparison}
\end{figure}

The collapse of the NHIM to
a single periodic trajectory
in dissipative regimes with temperature $T = 0$
observed in Sec.~\ref{sec:dissipative}
can be used to our advantage for temperatures above zero.
Even when noise is in play,
the fluctuating trajectories on the \ac{NHIM}
will approach a single fluctuating trajectory over long times,
which we will
refer to as the \textit{equilibrium trajectory} on the \ac{NHIM}.
It can be determined numerically
by propagating an arbitrary point on the \ac{NHIM} for a sufficiently long time
until the equilibrium is reached.
The initial build-up is discarded when calculating rates.
In the infinite time limit,
use of the equilibrium trajectory to obtain the average decay rate $\bar k$
eliminates its dependence on the initial
conditions $(y_0, {v_y}_0, t_0)$ in Eqs.~\eqref{eq:average} and
\eqref{eq:floquet-rate}.
That is, all contributions to the average decay rate
that would depend on the initial conditions
are dwarfed by the contributions of
the equilibrium trajectory.

Assuming that the long-term behavior is independent of the initial time $t_0$,
the equilibrium trajectory
can be used to construct the expected value
$\left<k\right>(T,\gamma)$ of
the reactant decay rate as
a function of the temperature $T$ and friction $\gamma$
\begin{equation}
    \left<k\right>(T,\gamma) = \bar k(y_0, {v_y}_0, t_0; T,\gamma)\,,
\end{equation}
which, due to the fact that the initial conditions will be forgotten by the
dynamics over time, is the same for any set of initial conditions
$(y_0, {v_y}_0, t_0)$.

\subsubsection{Instantaneous decay rates}
\label{sec:decay-over-time}

The influence of the noise on the average decay rate
is revealed by the time evolution of the instantaneous decay rate.
The rates shown in Fig.~\ref{fig:rate_comparison}(a) are plotted over
five saddle oscillations.
Despite the stochastic nature of the \ac{NHIM} at fixed orthogonal modes
$(y, v_y)$,
we find that the
instantaneous rate of the thermal trajectory on the \ac{NHIM} is smooth and
still
roughly follows the regular oscillation we find for zero temperature.
This can be attributed to the fact that the trajectories, as integrated
values over a noisy acceleration, have a smooth time evolution.
Despite the instant decay in the time correlation of the stochastic
force in the
fluctuation-dissipation relation
[Eq.~\eqref{eq:noise}], these trajectories have a finite
memory of their previous positions~\cite{zwan61a}.
However, over sufficiently long time scales, it is possible to obtain a time
evolution of the instantaneous reactant decay rate that resembles an
uncorrelated fluctuation, as can be seen in Fig.~\ref{fig:rate_comparison}(b).

\subsubsection{Temperature dependence of average decay rates}
\label{sec:temperature-dependence}

\begin{figure}
    \includegraphics[width=\linewidth]{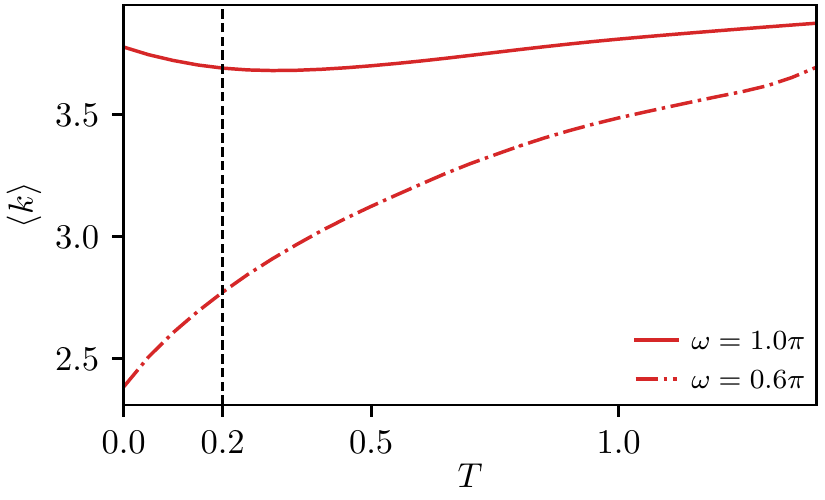}
    \caption{%
        Temperature dependent time-average of the instantaneous decay rate
        for the activated complex of the model system
        at two different driving frequencies, $\omega = 0.6 \pi$ and $1.0 \pi$.
        The dotted vertical line highlights the average rates at $T=0.2$
        corresponding to the data in
        Figs.~\ref{fig:rate_comparison} and~\ref{fig:floquet_nhim}.}
    \label{fig:average_over_kt}
\end{figure}

\begin{figure}
    \includegraphics[width=\linewidth]{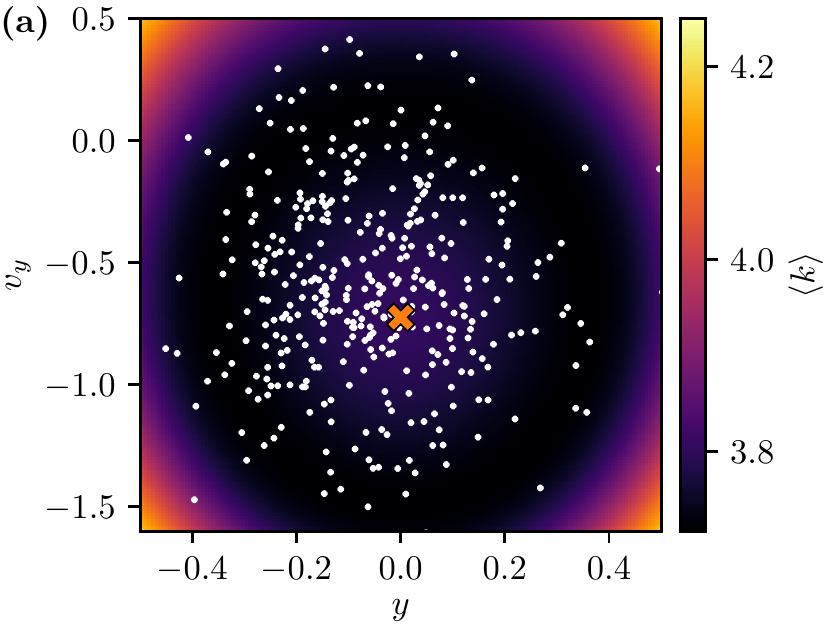}
    \includegraphics[width=\linewidth]{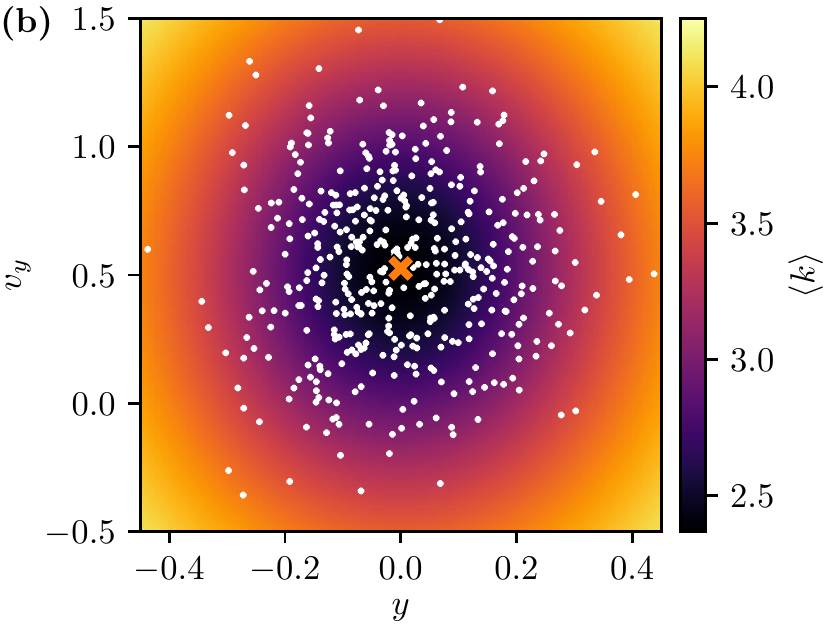}
    \caption{%
        Stroboscopic projections of a typical trajectory on
        the \acs{NHIM}
        for each of the saddle driving frequencies, (a) $\omega = \pi$
        and (b) $\omega=0.6\pi$,
        in the dissipated thermal case, $\gamma = 0.2$ and $T=0.2$,
        are shown as white filled circles.
        Refer to Fig.~\ref{fig:average_over_kt} for the corresponding
        average rates where the vertical dashed line crosses the curves.
        The color maps show the time-averaged decay rates $\left< k\right>$ of
        a non-thermal activated complex
        on the \acs{NHIM} for the initial conditions $(y,v_y)$ at time $t_0 = 0$.
        For reference, the fixed points of the stroboscopic maps
        in the non-thermal case are indicated by a cross.}
    \label{fig:floquet_nhim}
\end{figure}

The averaged rate
has to be determined over several hundreds---sometimes even a few thousand---%
saddle oscillations to obtain a statistically-sound converged average.
This is as a consequence of the need to achieve the limiting
equilibrium trajectory as observed in
Sec.~\ref{sec:decay-over-time}.
The resulting temperature dependence of the averaged decay rate for two sets of
parameters is shown in Fig.~\ref{fig:average_over_kt}.
These parameters were chosen, as they
give rise to two very different regimes in
the shape of the rate from concave to convex.
With a driving frequency of $\omega = 0.6\pi$, we obtain a
rate that is monotonically rising, as temperature rises. However, at a driving
frequency of $\omega = \pi$, we obtain a rate that at first decreases, before
it increases with rising temperature.

The change in behavior of the rate curves
may seem counter-intuitive at first.
One might expect that a higher temperature%
---\ie,\ a higher average energy---%
would cause the reaction to
surmount an energy barrier at a higher rate.
This expectation is not contradicted by the present results.
The decay rate computed here is the decay rate of the reactant population
close to the transition
state, \ie, the rate of reaction after the reactant has already
surmounted the energy barrier.
Such a rate does not include the increased population of
activated reactants that would arise from a higher temperature
and consequently does not have to increase accordingly.

We find that strong noise causes the activated complex
to fluctuate strongly near the \ac{NHIM}
as seen in Fig.~\ref{fig:floquet_nhim}.
When comparing these coordinate fluctuations with the phase space resolved
decay rates of the non-thermal system according to Fig.~\ref{fig:floquet_nhim}
or additional examples in the Supplemental Material~\cite{p22},
we can also see how a thermal activated complex
explores several transition states
of the non-thermal system.
This further suggests that the decay of the thermal
activated complex into reactants or products is related to the distribution
of the explored, non-thermal transition states.
This interpretation is consistent with the observation made in
Sec.~\ref{sec:rate_methods} that noise only affects the average decay rate by the
trajectory that is used to obtain it.
Even though
this analogy is not formally exact
since the
Jacobian $\boldsymbol{J}$ according to Eq.~\eqref{eq:jacobian}
contains the friction coefficient $\gamma$---%
which is not zero in the thermal case---%
the conjectured interpretation appears to hold for the
low-temperature
thermal case discussed in
Figs.~\ref{fig:average_over_kt} and~\ref{fig:floquet_nhim}.

A heuristic argument in support of the conjecture is as follows.
As temperature rises, the
activated complex deviates further from the
corresponding periodic trajectory at $T=0$.
This in turn causes the activated complex to explore transition states
that are further from said trajectory.
Moreover, the distribution of explored transition states expands
into a region with shrinking decay rates as temperature rises.
The equilibrium trajectory now resides in a local maximum
of decay rates, as is the case for $\omega=0.6\pi$,
and hence the decay of the activated complex decreases with rising
temperature.

\section{Summary and conclusion}

We have characterized the geometric structure of a model chemical
reaction, thereby taking into account both external driving and noise
and friction described by the Langevin terms.
We have shown that the temperature dependence of the activated complex
decay in this thermal system is linked to the distribution of the
phase space resolved decay rates on the \ac{NHIM} in the
non-dissipative case.
The decay rate of the activated complex depends on the external
driving and the temperature, and these dependencies can be used to
control the reaction.

In this paper we have investigated the thermal decay rates of
trajectories very close to the \ac{NHIM} based on equilibrium
trajectories located exactly on the \ac{NHIM}.
In future work it will be necessary to also study trajectories out of
the \ac{NHIM}.
An important question is whether
one can define a thermal equilibrium or at
least a stationary distribution on the \ac{DS} in these nonequilibrium
systems with which one can obtain the reaction rate.

Recently, we have investigated the influence of external driving on
decays in the geometry of the LiCN isomerization without considering
noise and friction~\cite{hern20m}.
Meanwhile, the dissipation arising from an argon bath
on that reaction was seen to be representable by the
Langevin terms~\cite{hern16c}.
Thus, the methods presented here open up the possibility
of considering the thermal effects
in a driven LiCN isomerization reaction
and other chemical reactions of interest.

\section{Acknowledgments}
The German portion of this collaborative work was supported
by Deutsche Forschungsgemeinschaft (DFG) through Grant
No.~MA1639/14-1.
RH's contribution to this work was supported by the National Science
Foundation (NSF) through Grant No.~CHE-1700749.
M.F.\ is grateful for support from the Landesgraduiertenf\"orderung of
the Land Baden-W\"urttemberg.
This collaboration has also benefited from support by the European
Union's Horizon 2020 Research and Innovation Program under the Marie
Skłodowska-Curie Grant Agreement No.~734557.

\bibliography{paper-q22}

\end{document}


\title{Supplemental Material for
``Thermal decay rate of an activated complex in a driven model chemical reaction''}

\author{Robin Bardakcioglu}
\author{Johannes Reiff}
\author{Matthias Feldmaier}
\author{J\"org Main}
\affiliation{%
Institut f\"ur Theoretische Physik 1,
Universit\"at Stuttgart,
70550 Stuttgart,
Germany}
\author{Rigoberto Hernandez}
\email[Correspondence to: ]{r.hernandez@jhu.edu}
\affiliation{%
Department of Chemistry,
Johns Hopkins University,
Baltimore, Maryland 21218, USA
}
\date{\today}
\maketitle
\acrodef{NHIM}{normally hyperbolic invariant manifold}

\begin{figure*}
    \centering
    \includegraphics[width=0.45\linewidth]{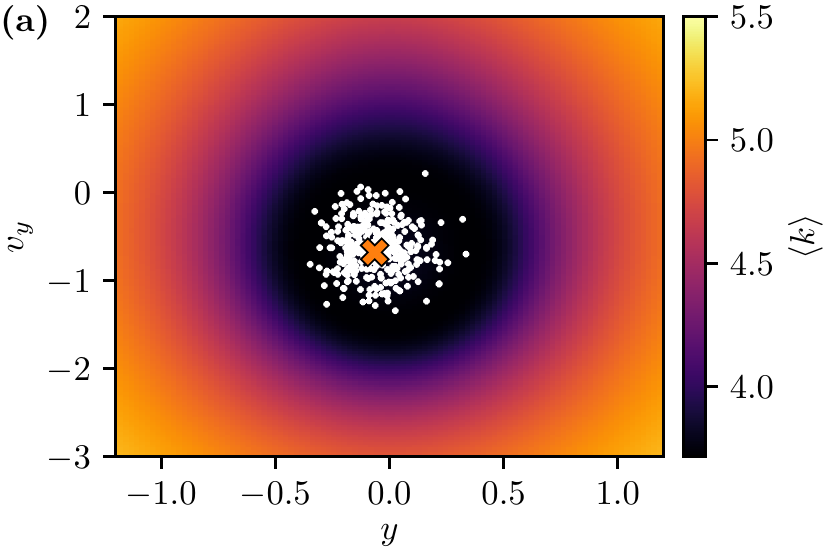}
    \includegraphics[width=0.45\linewidth]{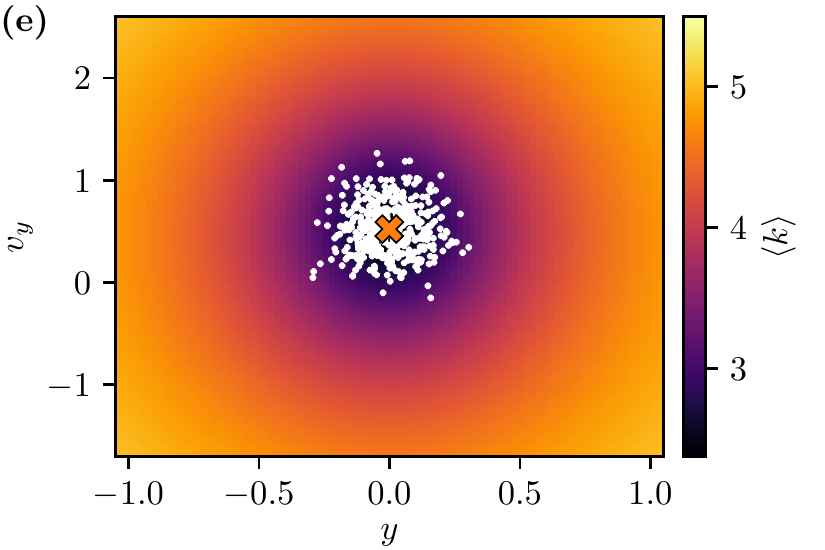}

    \includegraphics[width=0.45\linewidth]{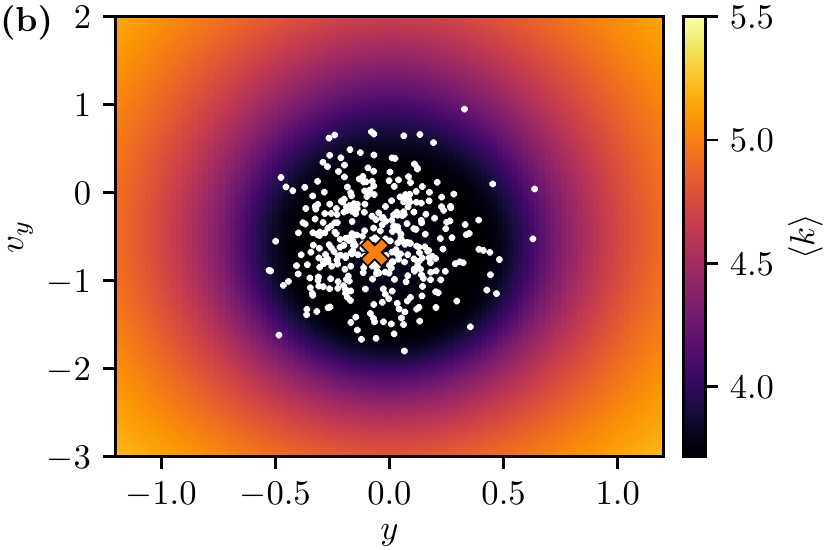}
    \includegraphics[width=0.45\linewidth]{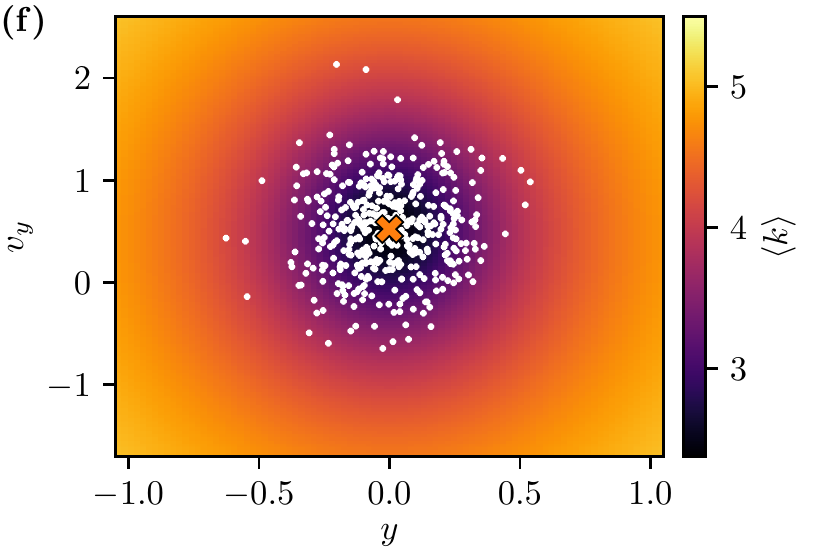}

    \includegraphics[width=0.45\linewidth]{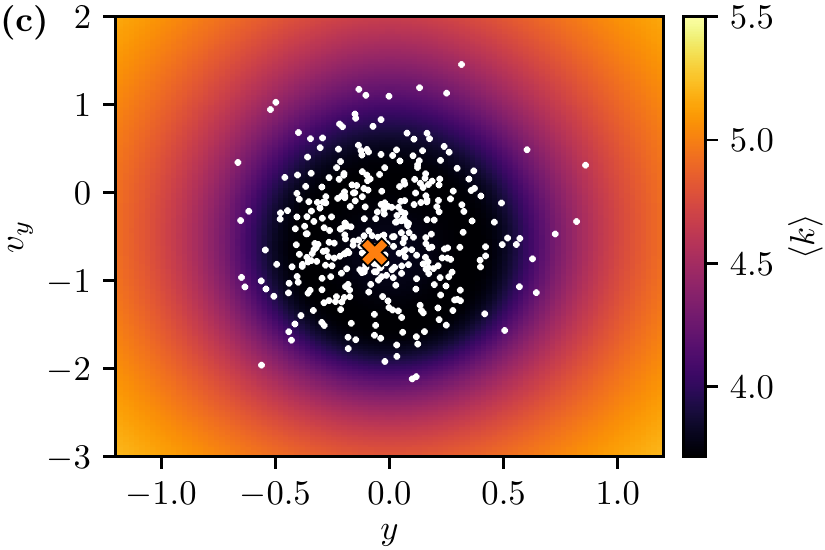}
    \includegraphics[width=0.45\linewidth]{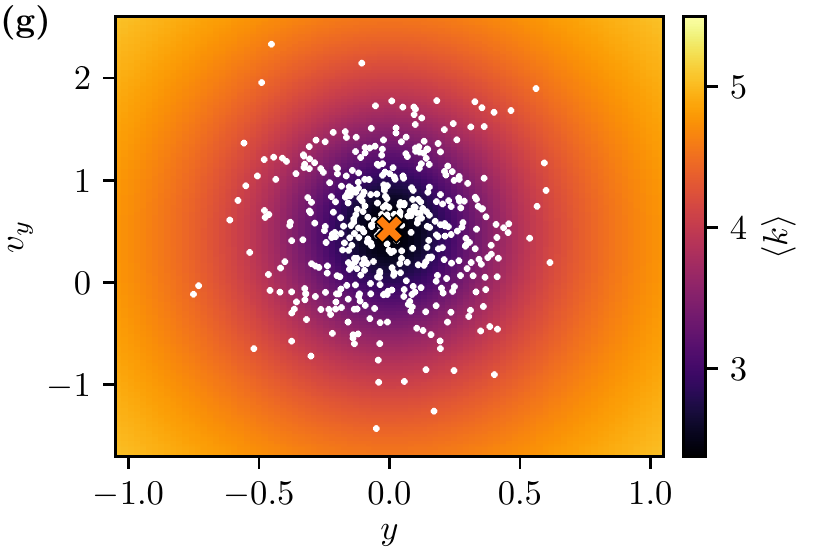}

    \includegraphics[width=0.45\linewidth]{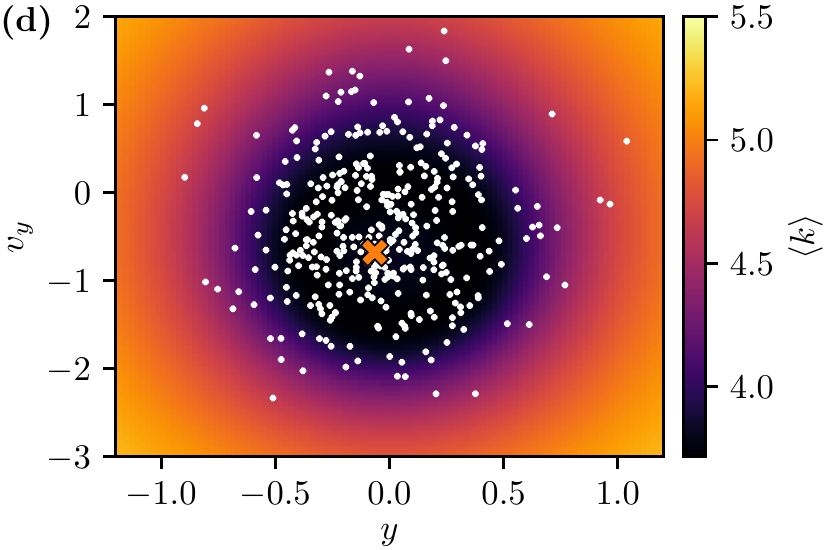}
    \includegraphics[width=0.45\linewidth]{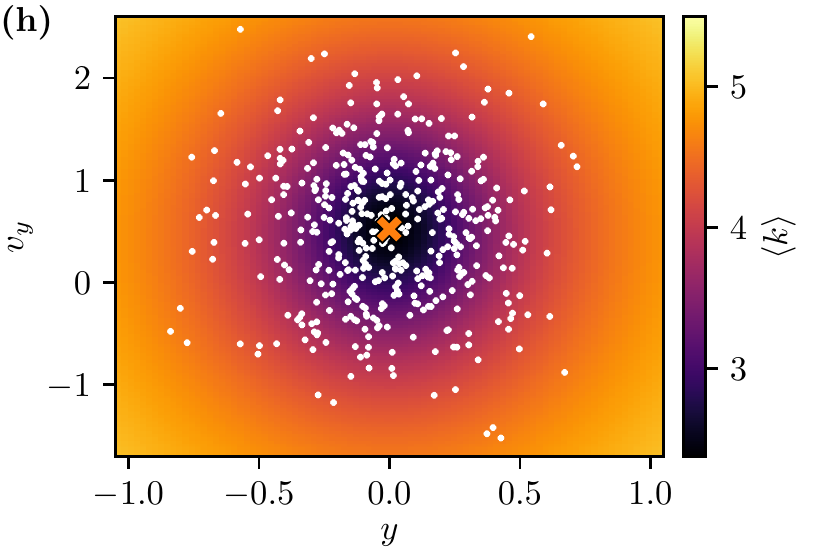}
    \caption{
        Stroboscopic view of a thermal trajectory
        at (a, e) $T=0.1$, (b, f) $T=0.3$, (c, g) $T=0.5$, and (d, h) $T=0.7$
        on the \acs{NHIM} for $\gamma = 0.2$ and
        oscillation frequency $\omega=\pi$ (a--d) and $\omega=0.6\pi$ (e--h).
        The color maps show the time-averaged decay rates $\ev{k}$ of
        a non-thermal activated complex
        on the \acs{NHIM} for the initial conditions $(y,v_y)$ at time $t_0 = 0$.
        The fixed point of the stroboscopic maps in the non-thermal case
        is indicated by a cross.}
    \label{fig:T}
\end{figure*}

This supplemental material includes the stroboscopic maps of the
thermal trajectories on the NHIM for $\gamma = 0.2$ oscillation
frequency $\omega=\pi$ (top) and $\omega=0.6\pi$ (bottom) at various
temperatures other than $T=0.2$ shown in Fig.~7 of the main text.
It serves to confirm that the data in at chosen temperature in the
main text is illustrative of the dynamics across a broad temperature range.
Specifically, Fig.~\ref{fig:T}(a)--(d) correspond to temperatures
equal to $T=0.1$, $0.3$, $0.5$ and $0.7$, respectively.
As claimed in the main text, the ``moat'' is visible in all of the
lower frequency driving cases.

